\documentclass[aps,prc,twocolumn,superscriptaddress,amsmath,showpacs,amssymb]{revtex4-1}
\usepackage{bm,physics}
\usepackage[T1]{fontenc}
\usepackage{textcomp}
\usepackage[dvips,final]{graphicx}
\usepackage{ulem, color}
\usepackage{hyperref}
\bmdefine{\ba}{a}
\bmdefine{\bb}{b}
\bmdefine{\bx}{x}
\bmdefine{\by}{y}
\bmdefine{\bz}{z}
\bmdefine{\bn}{n}
\bmdefine{\bp}{p}
\newcommand{\BM}{\begin{pmatrix}}
\newcommand{\EM}{\end{pmatrix}}

\begin{document}
\title {Existence of  higher nodal band states with 
   $\alpha$+$^{48}$Ca  cluster structure  in $^{52}$Ti \\
 }
%
\author{S.~Ohkubo}
\affiliation{Research Center for Nuclear Physics,
Osaka University, Ibaraki,
Osaka 567-0047, Japan}
\date{\today}
\begin{abstract}
\par
 It is shown  that  recently observed $\alpha$ cluster states a few MeV above the   $\alpha$ threshold energy in  $^{52}$Ti 
 correspond to  the     higher nodal band states with the $\alpha$+$^{48}$Ca
 cluster structure, i.e.  a vibrational mode in which 
 the intercluster   relative motion is excited.
    The existence of the higher nodal states in the $^{48}$Ca core region  in addition to the  well-known higher nodal states  in $^{20}$Ne and  $^{44}$Ti reinforces
the  importance of the concept of  vibrational motion due to clustering  even in  medium-weight  nuclei with a  $jj$-shell closed core.
The higher nodal band 
 and the shell-like ground band    in $^{52}$Ti   are described   in a unified way by a Luneburg lens-like deep local potential due to the Pauli principle, which  explains  the emergence of  backward angle anomaly (anomalous large angle  scattering (ALAS)) at low energies,  prerainbows   at intermediate energies  and nuclear rainbows  at  high energies in  $\alpha$+$^{48}$Ca scattering.  The existence of a $K=0^-$ $\alpha$ cluster band analog to $^{44}$Ti midway between the ground band and the higher nodal band is inevitably predicted.
 \end{abstract}


\maketitle

\par
 The $\alpha$  clustering is essential 
in the $0p$-shell and $sd$ shell region   and the nuclear  structure
has been  comprehensively understood from the $\alpha$ cluster viewpoint  \cite{Suppl1980}.  
In the $fp$ shell region, identification  of the higher nodal band states with the $\alpha$+$^{40}$Ca cluster structure  in the fusion excitation functions \cite{Michel1986A} lead to  the prediction  of a $K=0^-$ band, which is a parity-doublet partner of the ground band, in  the typical nucleus $^{44}$Ti \cite{Michel1986,Michel1988,Ohkubo1988}. The   observation of the  $K=0^-$ band   in experiment \cite{Yamaya1990B,Guazzoni1993} 
showed that the $\alpha$ cluster picture is also essential in $^{44}$Ti.  %
 Systematic theoretical and experimental studies in the $^{44}$Ti region \cite{Ohkubo1998,Michel1998,Yamaya1998,Sakuda1998,Ohkubo1998B,Fukada2009}  confirmed  the existence of the  $\alpha$ cluster  in the beginning of the $fp$-shell  above  the double magic nucleus $^{40}$Ca.
   
\par
  $\alpha$ clustering aspects in nuclei beyond $^{44}$Ti 
have been explored  in the medium weight  mass region  around A=50  such as $^{48}$Cr \cite{Descouvemont2002,Sakuda2002} and $^{46,50}$Cr  \cite{Souza2017,Mohr2017} as well as  in the  heavy mass region
such as  $^{94}$Mo and     $^{212}$Po 
      \cite{Ohkubo1995,Buck1995,Michel2000,Ohkubo2009}. 
 $^{52}$Ti, which  is a typical  nucleus with two protons and two neutrons outside the doubly closed core  $^{48}$Ca   analog to  $^{20}$Ne and $^{44}$Ti,  
 has been mostly studied in the shell model 
\cite{McGrory1967,Horie1973,Nakada1994,Janssens2002,Dinca2005,Zhu2009}.
     The ground band $0^+$, $2^+$ and $4^+$  states are  selectively enhanced  in the  $\alpha$-transfer reactions such as $^{48}$Ca($^{16}$O,$^{12}$C)$^{52}$Ti \cite{Faraggi1971} and $^{48}$Ca($^{12}$C, $^8$Be)$^{52}$Ti \cite{Mathiak1976}. 
However a   microscopic  $\alpha$+$^{48}$Ca cluster model calculation  with Brink-Boeker force B1  using the generator coordinate method (GCM) \cite{Langanke1982}
did not give the ground band  as well as in the   $\alpha$+$^{40}$Ca cluster model calculation  for $^{44}$Ti.
On the other hand, 
Ohkubo {\it et al.} \cite{Ohkubo1989A} and Ohkubo and Hiraoka\cite{Ohkubo1992}  reproduced  the ground  band   of $^{52}$Ti  in the $\alpha$ cluster model     with a  local potential similar to $^{44}$Ti \cite{Michel1986,Michel1988}.
    No  experimental data   that suggest  clear  $\alpha$ cluster states 
        hampered    to conclude that the $\alpha$ cluster picture persists in $^{52}$Ti  in the  jj-shell closed $^{48}$Ca core region.

\par
Very recently Bailey {\it et al.} \cite{Bailey2019} reported
  that they newly observed $\alpha$ cluster states at the highly excited energy region in $^{52}$Ti. This prompts us
to clarify  the nature of the observed $\alpha$ cluster states theoretically,  especially  to which band they belong.
In this respect I note that the emergence of a  cluster structure is a
 consequence of the Pauli principle\cite{Ohkubo2016},  which causes a  Luneburg lens-like deep intercluster potential that accommodates  the Pauli-allowed cluster states in the low energy region    and   a nuclear rainbow due to astigmatism of the lens with a diffuse surface  at high energies\cite{Michel2002,Ohkubo2016}.
 The cluster structure  and the nuclear rainbow  are  the two  aspects of the phenomena caused by the same Luneburg lens-like  inter-nucleus interaction  \cite{Ohkubo2016}. Therefore
   it seems useful to clarify   the nature of the observed $\alpha$ cluster states in $^{52}$Ti from the viewpoint of understanding  the ground band states and scattering phenomena for  $\alpha$+$^{48}$Ca including nuclear rainbows   in a 
unified way. 
 \par
 The purpose of this paper is to show   that the newly observed three $\alpha$ cluster states correspond to  the higher nodal
 states with the  $\alpha$+$^{48}$Ca cluster structure in $^{52}$Ti by studying the  nuclear rainbows at high energies, the Airy structure of the prerainbows at intermediate energies, the backward angle anomaly (anomalous large angles  scattering (ALAS))  at  low energies, the $\alpha$ cluster structure near the $\alpha$ threshold energy, and the ground state band simultaneously.
 The existence of a higher nodal excitation mode
   inevitably predicts the existence of  a $K=0^-$ band state with the 
 $\alpha$ cluster structure midway between the higher nodal band and the ground band in $^{52}$Ti.

\begin{figure}[t!]
\includegraphics[width=8.6cm]{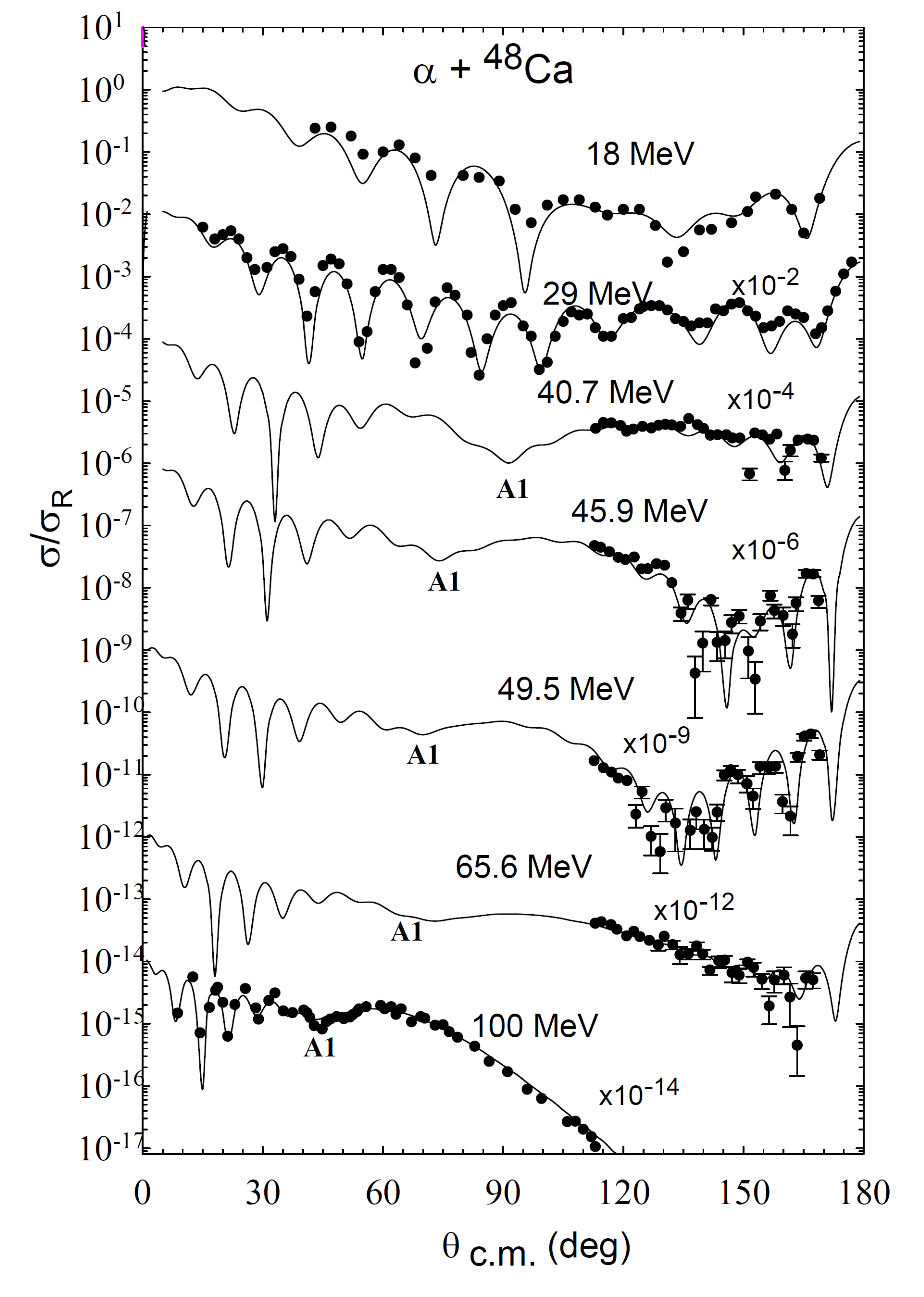}
 \protect\caption {The  angular distributions  of cross sections (ratio to Rutherford scattering) in   $\alpha$  particle scattering from $^{48}$Ca  calculated with the optical model potentials in Table I (solid lines) are compared with the experimental data (points) taken from Refs.\cite{Gaul1969,Stock1972,Eickhoff1975}.
 $A1$    indicates  the Airy minimum.
 }
\end{figure}

The anomalous rise of cross sections at backward angles in $\alpha$ particle scattering, ALAS, which was first typically found in $\alpha$ particle scattering from $^{40}$Ca \cite{Budzanowski1966,Gruhn1966},  is seen  persistently in the scattering  from the closed nucleus $^{48}$Ca \cite{Gaul1969} at low energies, $E_L$=18-29 MeV.
 Stock {\it et a1.} \cite{Stock1972} extended the measurement at backward
angles to intermediate energies at $E_L$=40.7-65.6 MeV where prerainbows appear. The nuclear rainbow 
was observed at high energies above $E_L$=100 MeV  \cite{Eickhoff1975,Gils1980}.  
 I  use a Woods-Saxon squared local  potential, which can simulate a Luneburg lens well
 as in the case of  the $\alpha$ cluster study in $^{44}$Ti \cite{Michel1986,Michel1988}. In the optical model analysis  imaginary potentials  with a  Woods-Saxon 
 and its derivative 
  form factors are introduced, $U(r) = - V f^2(r;R_v,a_v)+ V_{coul}(r) -i W f(r;R_w,a_w) -i 4a_s W_s\frac{d}{dr}f(r,R_s,a_s)$
\noindent with $f(r;R,a) =1/\{1+\exp[(r-R)/a]\}$. The Coulomb potential is assumed to be a uniformly charged sphere with a reduced radius  $r_{c}$=1.3 fm.

First  I analyze the experimental angular distributions in $\alpha$+$^{48}$Ca scattering  with the optical potential model.
For the real part of the optical potential, I started from the unique  potential used in the  systematic analysis of $\alpha$+$^{40}$Ca scattering   over a wide range of incident energies 
 \cite{Delbar1978} and the  $\alpha$ cluster structure study of  $^{44}$Ti \cite{Michel1986,Michel1988}. 
  The obtained  potential parameters that fit the experimental angular distributions  are  listed in Table~I.  
At the lower energies below $E_L$=29 MeV the characteristic behavior of ALAS angular distributions rising toward the extreme backward angles, which  
is difficult to reproduce using the average optical potential 
as noted in Ref.\cite{Gaul1969},   is  slightly  seen. 
The imaginary potential parameters are searched to fit the data.
Different from  $\alpha$+$^{40,44}$Ca  \cite{Delbar1978}, a surface absorption was  needed to reproduce the angular distributions at lower energies, which  seems to be  due to the effect of the extra neutrons in the surface region of $^{48}$Ca.
Above  $E_L$=40 MeV  no surface absorption was needed in the analysis.  For the real potential the radius parameter $r_v$ is adjusted  around 1.35 fm with a fixed $a_v$=1.29.  For the imaginary potential  $r_w$  and   $a_w$ are fixed at 1.25  and 1.00 fm, respectively,  except slight modifications of $r_w$   at $E_L$=100 MeV.
 In Fig.~l the calculated results are compared with the experimental data.
 The calculations
reproduce the characteristic feature of the experimental angular distributions  well  up to the backward   angles.  Although the experimental data are not available in  the forward and intermediate  angle regions at  $E_L$=40.7-65.6 MeV,  the  data at  110-140$^\circ$, which determine the slope of the  fall-off  in the angular distributions  of the prerainbows,   are sensitive enough to   constraint  the real part of the potential.  The calculations reproduce the slope of the prerainbows and the characteristic oscillations of the experimental angular distributions.  
At $E_L$=100 MeV the nuclear rainbow  scattering with the lit side minimum  at around   $\theta$=42$^\circ$ is  reproduced well. The minimum,  beyond which the fall-off of the  angular distribution in the darkside of the rainbow follows,  corresponds to the first Airy minimum $A1$, which is  clearly seen in  the  angular distribution  calculated by switching off  the imaginary potential. 
As seen  in Fig.~1, this $A1$  evolves from the Airy minimum $A1$   at around 90$^\circ$  at $E_L$=40.9 MeV.
The evolution 
  is  similar to   $\alpha$+$^{40}$Ca scattering in Ref\cite{Delbar1978}. 
 
\begin{table}[t!]
\begin{center}
\protect\caption{The optical potential parameters used in Fig.~1 and the volume integrals per nucleon pair, $J_v$, in unit of MeVfm$^3$ for the real potentials. 
$E_L$, $V$, $W$ and $W_s$ are in units of MeV and $r_v$, $a_v$, $r_w$, $a_w$, 
$r_s$ and  $a_s$ are in units of fm.}
\begin{tabular}{ccccccccccc}
 \hline
  \hline
  $E_L$  &    $J_v$      &   $V$ & $r_v$ & $a_v$ &     $W$ &$r_w$& $a_w$& $W_s$& $r_s$ & $a_s$  \\
  \hline
  18   &     380    &  192.6 & 1.38 & 1.29    &  28.1 &1.00 &0.047 
       &   6.3&1.31&0.265\\
    29   &     342    &     189 & 1.34 & 1.25  &      28.3&1.02&0.975    
            &4.9&1.41&0.227\\ 
 40.7   &    355    &  180 & 1.38 & 1.29    &  23&1.25&1.0  &  &  & \\ 
  45.9   &     348   & 180  & 1.37 & 1.29  &  24 &1.25&1.0  &  &  & \\ 
 49.5   &     355    & 180  & 1.38 & 1.29  &  26&1.25&1.0 & &  & \\ 
 65.6 &   \ {296}    & 160  & 1.35 & 1.29  &  29 &1.25&1.0   &  &  &\\ 
 100   &     292    &  164 & 1.33 & 1.29  &  28.5 &1.31 &1.0 &  &  &\\ 
 \hline
 \hline
\end{tabular}
\end{center}
\end{table}

\begin{figure}[t!]
\includegraphics[width=5.5cm]{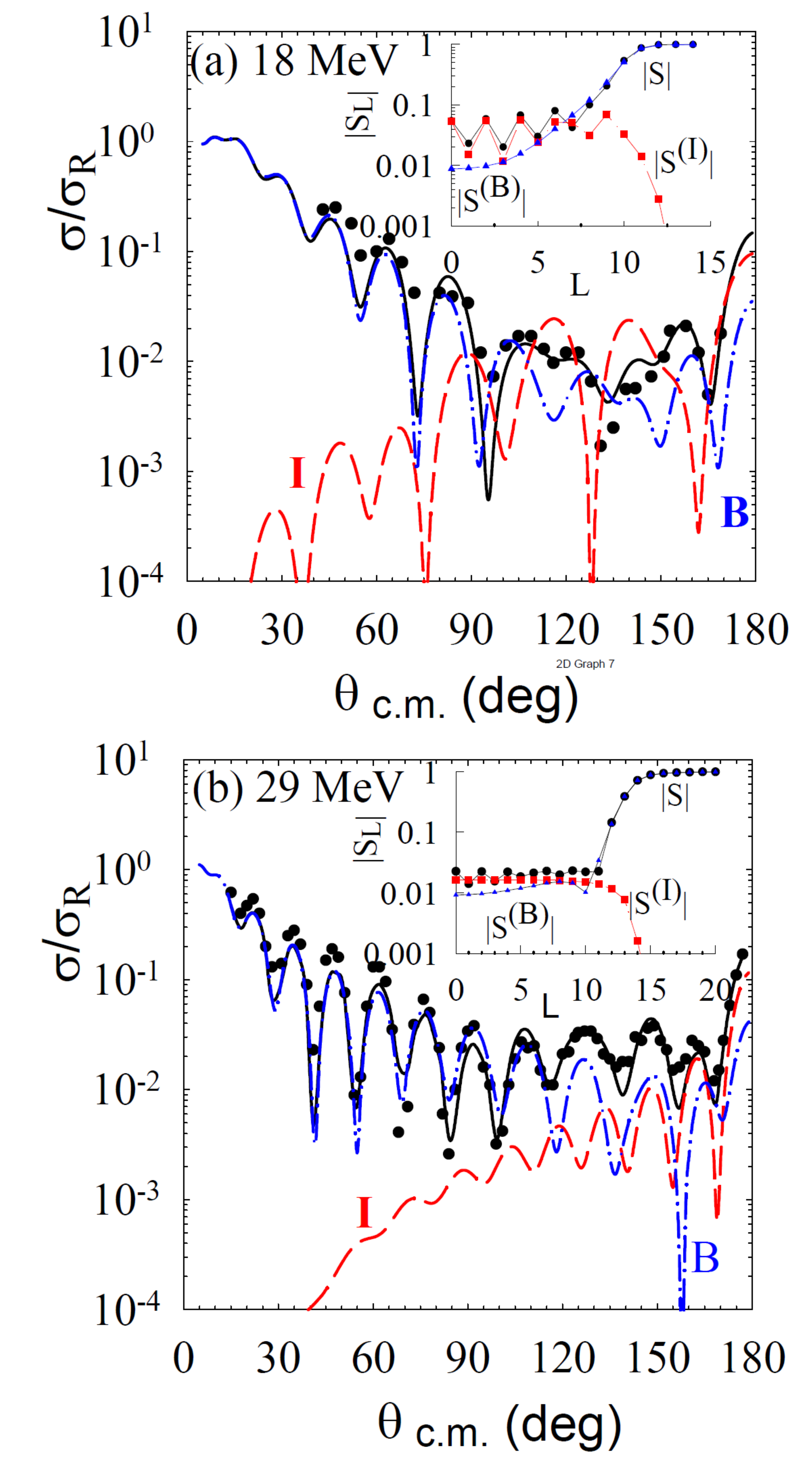}
 \protect\caption{(Color online)
The calculated  angular distributions  of cross sections (ratio to Rutherford scattering) in   $\alpha$  particle scattering from $^{48}$Ca (solid lines) at (a) $E_L$=18  and (b) 29 MeV     in Fig.~1   are decomposed into the internal-wave (dashed lines) and the barrier-wave (dash-dotted lines) contributions. The points are the  experimental data from Ref.\cite{Gaul1969}.
   In the insets,  the  reflection coefficient  ($|S_L|$) is  decomposed into   the internal-wave ($|S_L^{(I)}|$)  and   the barrier-wave ($|S_L^{(B)}|$).  The lines are to guide the eye.
 }
\end{figure}

\par
 The appearance of the nuclear rainbow and the Airy structure in the prerainbows, which  are sensitive to the internal region of the real part of the potential, shows that the scattering is not strongly absorptive. In the lower energies at 29 and 18 MeV where the ALAS appears as  a precursor of the prerainbows,   scattering  is also
 sensitive to  the internal region of the potential.  In Fig.~2 the calculated angular distributions at $E_L$=18 and 29 MeV are decomposed using the technique of Ref.\cite{Albinski1982} into the barrier-wave component reflected at the surface and the internal-wave component, which penetrates deep  into the internal region of the potential,    are displayed. 
In the last peak of the angular distribution toward 180$^\circ$,    the internal-wave contribution is dominated. 
However, the interference of the two components, which is seen only in the intermediate angles   in the $\alpha$+$^{40}$Ca system, occurs even  at the  backward angles.
The characteristic behavior of the experimental angular distribution beyond  $\theta$=90$^\circ$ is well reproduced by the interference between the internal-wave and the barrier-wave contributions. 
In the insets, one  sees that at 18 MeV the reflection coefficient of  the internal-wave is considerably   larger than  that of  the barrier-wave for the low partial waves, which shows that scattering is  transparent.

\begin{table*}[t!]
\begin{center}
\protect\caption{ (Color online) The calculated energy with respect to the $\alpha$ threshold  $E$,   excitation energy $E_x$, intercluster rms radii $<R^2>^{1/2}$ and  $B(E2)$ values in unit of $e^2$fm$^4$ for the $J$ $\rightarrow$ $J-2$ transitions for the $N=12$ and $N=13$ band states in $^{52}$Ti. Theoretical  $B(E2)$ values are compared with the experimental data \cite{Goldkuhle2019} and the shell model calculations \cite{Nakada1994}.
}
\begin{tabular}{rcccccccrccc}
 \hline
  \hline
      \multicolumn{7}{c}{$N=12$}  & \multicolumn{5}{c}{ $N=13$}  \\
  $J$   & $E$ & $E_x$&  $<R^2>^{1/2}$&  \multicolumn{3}{c}{$B(E2)$  ($J$ $\rightarrow$ $J-2$)}    
   &     $J$ & $E$ &$E_x$& $<R^2>^{1/2}$ & $B(E2)$ ($J$ $\rightarrow$ $J-2$) \\
    &    (MeV) &   (MeV)&  (fm)&exp.\cite{Goldkuhle2019}& This work&  Ref.\cite{Nakada1994} & &(MeV)  &(MeV) &(fm) & This work \\
   \hline
  $0^+$     &  -7.66  &0.0   &4.57     &  &     &     &$1^-$ & -0.76& 6.90& 5.05\\ 
  $2^+$     &  -7.36 	&0.31&4.53 &86$_{-4}^{+5}$ &108 & 100      &$3^-$ & -0.18& 7.48&4.99&212 \\  
  $4^+$     &  -6.73 	&0.93&4.50&  109$_{-13}^{+16}$ &142 & 134      &$5^-$ & 0.79&8.45& 4.89&223 \\ 
  $6^+$     &  -5.83 	&1.83&4.39& $100_{-6}^{+7}$  &134   & 88.6   &$7^-$ &2.14& 9.80&4.74&196 \\ 
  $8^+$  &  -4.67 &2.99&4.28 &8.8$_{-1}^{+1}$  &109&   &$9^-$ & 3.82&11.48& 4.56&152 \\
     $10^+$   &  -3.27 &4.39	&4.16    &  &76    &   &$11^-$ & 5.82 &13.48& 4.37&100 \\  
\hline
 \hline
\end{tabular}
\end{center}
\end{table*}

\par
The real part of the optical potential
   obtained in the analysis of $\alpha$ particle scattering is useful for the $\alpha$-cluster structure study in $^{52}$Ti. 
 The strength of the real part of the optical potentials is known to have
      energy dependence to decrease toward  the threshold (threshold anomaly) \cite{Mahaux1986} and the strength must be adjusted  in the $\alpha$ cluster calculations. 
In fact, the lowest Pauli-allowed state obtained   using the potential at $E_L$= 18 MeV (29  MeV) is overbinding compared  with the experimental value of  -7.76 MeV. 
%
In Fig 3 the  energy levels calculated in the bound state approximation  using the potential  at 18 MeV   in Table I  with the adjusted strength $V$=166.8 MeV,  which is   tuned to reproduce the binding energy of the ground state,    are shown. The  states with $N= 2n$+$L$ $<$12 and $L=12$ with $N=12$ are forbidden by the Pauli principle where $n$ and $L$ are   the number of nodes in the wave functions and the orbital angular momentum of the relative motion, respectively. 
The calculated $J_v$=329 MeVfm$^3$ of the potential is comparable to 350 MeVfm$^3$ for the $\alpha$+$^{40}$Ca system \cite{Michel1988}.
 The calculated ground band states fall well  in   correspondence   to  the experimental ground band. The observed higher spin states show antistretching  deviating  from a   rotational-like spectrum similar to  $^{20}$Ne \cite{Michel1989} and  $^{44}$Ti, which    can be reproduced by    taking into account the  $L$-dependence of the potential, as  discussed 
   in the $\alpha$ cluster structure in $^{20}$Ne \cite{Michel1989}, $^{44}$Ti \cite{Michel1986,Michel1988},  $^{94}$Mo \cite{Ohkubo1995,Souza2015}, $^{212}$Po \cite{Ohkubo1995,Ni2011} and  recently in $^{46,50}$Cr  \cite{Mohr2017}. 

 \begin{figure}[t!]
\includegraphics[width=7.6cm]{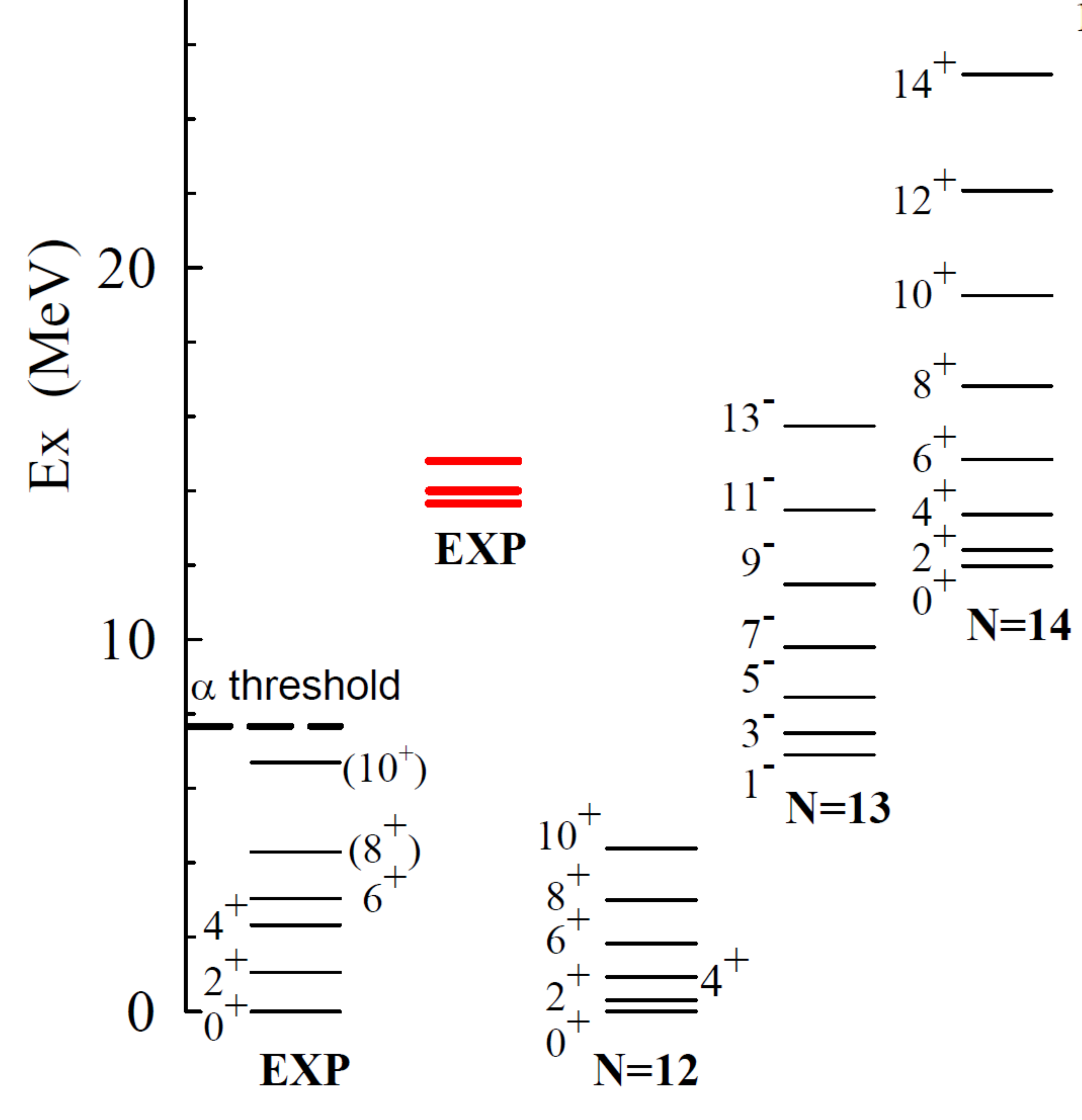}
 \protect\caption{Calculated energy levels of $^{52}$Ti are compared with the  experimental ground  band \cite{Dong2015} and the newly observed three  excited $\alpha$ cluster states (red thick solid lines) \cite{Bailey2019}.
}
 \end{figure}
\par
 In Table II one  sees that the calculated $B(E2)$ values of the ground band are considerably large
compared with  the single particle unit 11.5 $e^2$fm$^4$   and in agreement with the observed values without effective charges. 
In the shell model calculations in the fp-shell nuclei, usually large effective charges, such as   $e_\pi$=1.3-1.5 and $e_\nu$=0.6-0.8 in Ref.\cite{Nakada1994}
 are needed to reproduce the experimental $B(E2)$ values.  
The large $B(E2)$ values come from the collectivity due to the $\alpha$-clustering  and  one of the origins of the large effective charge may be ascribed to the $\alpha$-clustering degree of freedom.
 The rms distance between the $\alpha$ particle and the $^{48}$Ca core of the ground band is considerably smaller than 5.15 fm, 
 the sum of the rms  charge  radii of the  free $\alpha$ (1.676 fm) and $^{48}$Ca (3.477 fm) \cite{Angeli2013}. The two clusters overlap significantly.
 
\begin{table}[bth]
\begin{center}
\protect\caption{ Resonance energies and widths for the  $N=14$  band states in $^{52}$Ti, together with the corresponding  dimensionless reduced widths $\theta^2$ calculated with channel radii  $a$=7.5 and 8 fm. 
}
\begin{tabular}{ccccc}
 \hline
  \hline
  $J$ &    $E_{res}$  &  $\Gamma_{L}$ & \multicolumn{2}{c}{$\theta^2_L (\%)$} \\
     &  (MeV)  & (keV)&$a=7.5$  & $a=8.0$ \\ 
   \hline
  $0^+$   & 4.66  &   33   & 78    & 46  \\ 
  $2^+$   & 5.12  &   62   & 100    & 59  \\ 
  $4^+$   & 6.16  &   60   & 56    & 33 \\ 
   $6^+$  & 7.77  &   74   & 45    & 25 \\ 
     $8^+$  & 9.92  &    47   & 22    & 12 \\ 
      $10^+$ & 12.55  &   29   & 12    & 6 \\ 
      $12^+$ & 15.60  &   7   & 3    & 1 \\  
     $14^+$ & 18.98  &   <1   & <1   & <1 \\
      \hline
 \hline
\end{tabular}
\end{center}
\end{table}

\par
In Fig.~3 the calculation predicts the $N=$14 $\alpha$ cluster band above the $\alpha$ threshold energy.  One sees that  the newly  observed  three $\alpha$ cluster states 
  in Ref.\cite{Bailey2019} correspond in excitation energy  to the $0^+$, $2^+$ and $4^+$ states of the $N=14$ band. The energy intervals among the three states also  correspond to the calculation well. In fact, the ratio $R=(E_x(4^+)-E_x(0^+))/(E_x(2^+)-E_x(0^+))\simeq3.4$ for the observed three states, which shows that they can be considered to form a  rotational band, agrees well with   $R=3. 2$ of the theoretical  $N=14$ band.  Also the estimated rotational constant $k\simeq$57 keV for the observed states is close to the theoretical  $k=$69 keV for the $N=14$ band where $k\equiv\hbar^2/2\cal{J}$ with $\cal{J}$ being the moment of inertia. Here it is to be noted that the four  states  observed  in $^{44}$Ti \cite{Bailey2019} using the same technique as in $^{52}$Ti   also correspond to the $0^+$, $2^+$,  $4^+$ and $6^+$ states of the $N=14$ band \cite{Ohkubo1998B,Michel1998} well. 
The identification of the   $N$=14 higher nodal  band a few MeV above the $\alpha$ threshold energy 
 as an analog band observed  in $^{44}$Ti, in which  relative motion  between $\alpha$ and $^{48}$Ca is  one more excited compared with the ground band,   
  gives strong support to $\alpha$ clustering in $^{52}$Ti. 
 The  intercluster  rms radii  of the $N=$14 band  calculated in the bound state approximation,   5.99, 5.95, 5.84, 5.65 and 5.36 fm for the  $0^+$, $2^+$,  $4^+$,  $6^+$,  and  $8^+$  states, respectively,  
  are larger than the sum of those of the free $\alpha$ and $^{48}$Ca nuclei, which shows that this band has a well-developed $\alpha$ cluster structure.
The degree of $\alpha$ clustering is more clearly seen in the considerably large dimensionless reduced widths $\theta^2_L$ in Table III, which are calculated  from the $\alpha$ decay width $\Gamma_{L}$ at the resonance energy $E_{res}$ using the formula  $\Gamma_{L}=2 P_L(a) \gamma^2_L(a)$, $\gamma^2_L (a)=\theta^2_L (a) \gamma^2_w(a)$ and  $\gamma^2_w(a)=3\hbar^2/2\mu a$ with $P_L(a)$, $\gamma^2_L(a)$ and $\gamma^2_w(a)$ being the penetration factor,  reduced width and the Wigner limit value  at  a channel radius $a$, respectively. $\mu$ is the reduced mass.
 $\Gamma_{L}$ is calculated  from the phase shift $\delta_L$ using $\Gamma_{L}=2/(\frac{\partial{\delta_L}}{\partial{ E_{c.m.}}})$$_{E_{c.m.}=E_{res}}$.
 
\par  
 In Fig.~3 the  calculation inevitably locates the $N=13$  $K=0^-$ band with  the $\alpha$+$^{48}$Ca structure midway between  the ground band ($N=12$) and the higher nodal band ($N=14$).  The  $K=0^-$ band, which starts near the $\alpha$-threshold, is a parity-doublet partner of the ground band. The existence of such a  $N=13$ $K=0^-$ band  between the ground band and the higher nodal band has been already confirmed experimentally   in  $^{44}$Ti 
 \cite{Yamaya1990B,Guazzoni1993,Yamaya1998,Fukada2009}. 
  The  calculated  intercluster distances of the band states in Table II are slightly smaller than those calculated for the $N=13$ band in $^{44}$Ti \cite{Michel1986}. 
This suggests the $\alpha$ clustering of this band is smaller than that in  $^{44}$Ti.
I confirmed  that almost the same band structure as in Fig.~3 is obtained in  the calculations using  the potential at $E_L$=29 MeV in Table I with  the strength $V$  adjusted to reproduce the binding energy of the ground state.
The experimental observation of the member states of the  $N=13$ band would give further  support to $\alpha$ clustering  with the  parity-doublet structure  in the $^{48}$Ca core region.

 \begin{figure}[t]
\includegraphics[width=6.5cm]{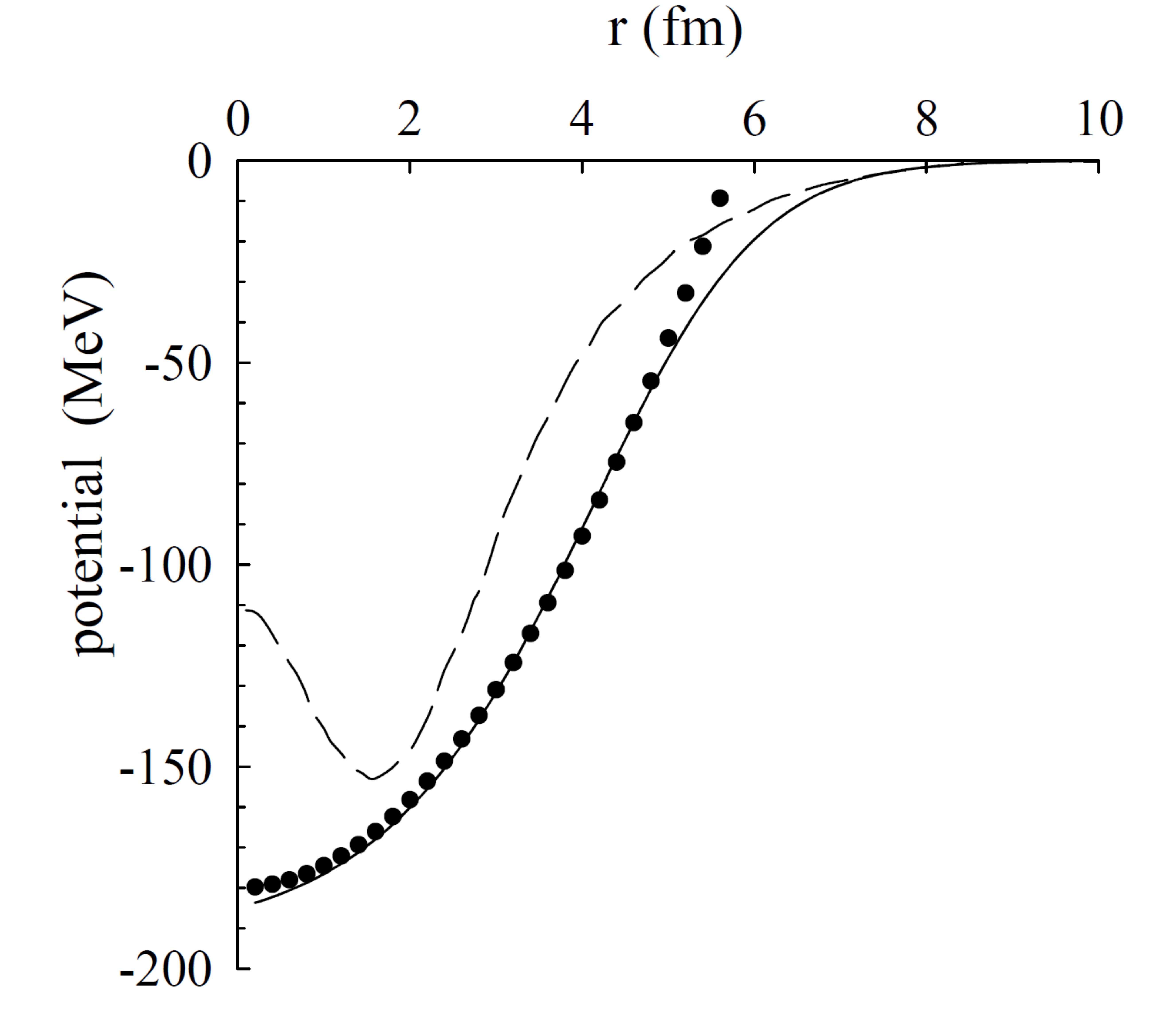}
 \protect\caption{The $\alpha$-$^{48}$Ca potential at $E_L=$18 MeV (solid line) is  compared with the Luneburg lens potential with $R_0$=5.75 fm  and $V_0$=180 MeV (points). The energy-independent equivalent local potential (for $L=$0) in  the GCM calculation from Ref.\cite{Wintgen1983} is shown by the dashed line. 
 }
 \end{figure}

In Fig.~4 the  potential at $E_L$=18 MeV is displayed in comparison with a Luneburg lens \cite{Luneburg1964} potential,
 which   decreases  radially from the  center to the outer surface $r=R_0$ and refracts  all the parallel  incident  trajectories  to the focus $r=R_f$  ($< R_0$).
The Luneburg lens potential  is  a truncated harmonic oscillator potential \cite{Michel2002} given  by 
$V(r) = V_0 \left(  {r^2}/{R_0^2}-1 \right)$ for $r \leq R_0$ and  $V(r) = 0$ for 
$r > R_0$.
One sees  that in the internal region  $r<5$ fm the potential 
 resembles the Luneburg lens potential. This is the reason why the present potential that
embeds the Pauli-forbidden states with $N<$12 deeply in the potential
 locates the $N=12$ and $N=14$  cluster band states in correspondence to experiment and predicts the unobserved $N=13$  band. The diffuse surface  of the potential at  $r>5$ fm where deviation from  the Luneburg lens is clear  causes  astigmatism of the lens, i.e.,  nuclear rainbow at high energies.
The  volume integral per 
pair nucleon pair, $J_v$=342 MeVfm$^3$ at  29 MeV, is as large as      $J_v$=345 MeVfm$^3$ of the potential by   Michel and Vanderpoorten \cite{Michel1979}
   obtained in the model-independent analysis of the angular distribution at $E_L$=29  MeV.
  The volume integral is also consistent with the value of the global potential for the  $\alpha$+$^{40}$Ca system, 350 MeVfm$^3$  \cite{Michel1979} at the same energy.
One finds that, in Fig.~4,  the equivalent local 
potential \cite{Wintgen1983} of the  microscopic GCM cluster model calculation\cite{Langanke1982}    belongs to a shallower potential family, which is unable to describe the nuclear rainbow. This explains why     the lowest Pauli-allowed  band with $N=$12 corresponding to the ground band  does not appear below the $\alpha$ threshold energy in Ref. \cite{Langanke1982}.

\par
To summarize,  the newly observed three $\alpha$ cluster states are found to correspond to the $N=14$ higher nodal  band states, the $0^+$, $2^+$ and $4^+$ states, which are the nodal excited states of the relative motion of the $\alpha$+$^{48}$Ca cluster structure in $^{52}$Ti. This gives  strong support to the persistence of the $\alpha$ cluster structure in $^{52}$Ti. The calculated lowest Pauli-allowed $N=12$ band is found to  correspond well to the experimental  ground band and the large experimental $B(E2)$ values are reproduced  without effective charge.
The local potential that describes backward angle anomaly (anomalous large angle scattering), prerainbows and nuclear rainbow in a wide range of  incident energies in $\alpha$+$^{48}$Ca scattering, the $N=14$ and $N=12$ $\alpha$ cluster bands, predicts inevitably the existence of a $K=0^-$ band ($N=13$), which is a parity-doublet partner of the  ground band, near the $\alpha$ threshold midway between the ground band and the $N=14$ higher nodal band.  Observation of the $K=0^-$ band  states  would give  further support to $\alpha$ clustering in the $jj$-shell closed $^{48}$Ca region.  

 \par
 The author  thanks the Yukawa Institute for Theoretical Physics, Kyoto University for    hospitality extended  during  a stay in  2019.

\end{document}